\documentstyle[twocolumn,prl,aps,epsf]{revtex}

\begin{document}

\newcommand {\be} {\begin{equation}}
\def \rem #1 {({\sl [#1]})}

\newcommand{\HH}{{\cal H}}
\newcommand{\Hop}{{\hat{\HH}}}
\newcommand{\Hsel}{{\Hop_{sel}}}
\newcommand{\Hbiq}{{\Hop_{biq}}}
\newcommand{\Hinf}{{\Hop_{inf}}}
\newcommand{\Ubiq}{{U_{biq}}}
\newcommand{\tCbiq}{\tilde{C}_{biq}}
\newcommand{\Eal}{{E_{inf}}}

\newcommand {\rr} {{\bf r}}
\newcommand {\ee} {{\hat{\bf e}}}
\newcommand {\mm} {{\hat{\bf m}}}
\newcommand {\nn} {{\hat{\bf n}}}
\newcommand {\nnmu} {{{\bf n}_\mu}}
\renewcommand {\ss} {{\bf s}}
\renewcommand {\SS} {{\bf S}}
\newcommand \ST{{\SS_{tot}}}
\newcommand \LC{{\cal{L}}}
\newcommand{\RC}{{\cal R}}

\newcommand {\tJ} {{\tilde{J}}}
\newcommand {\tK} {{\tilde{K}}}
\newcommand {\fourspin}{four-spin~}
\newcommand {\BS}{BS~}
\newcommand {\mmstate}[1]{\{ \mm_\alpha #1\}}
\newcommand {\Sall}{S}
\newcommand {\Ssub}{\tilde{S}}
\newcommand {\Nall}{N}
\newcommand {\Nsub}{\tilde{N}}
\newcommand {\Psing}{{\hat {\cal P}_s}}   
\newcommand {\ttot}{t^{tot}}


\twocolumn[\hsize\textwidth\columnwidth\hsize\csname@twocolumnfalse%
\endcsname

\draft

\title{Semiclassical eigenstates
of four-sublattice antiferromagnets}

\author{Christopher L. Henley and Nai-gong Zhang}

\address{Dept. of Physics, Cornell University,
Ithaca NY 14853-2501}
\maketitle

%
\begin{abstract}
Applying Bohr-Sommerfeld quantization and the topological phase of
spin path integrals, one can determine the multiplicities, 
lattice symmetries, and eigenvalue clustering pattern of the 
low-lying singlet eigenstates of the triangular and fcc antiferromagnets 
with 4-sublattice classical ground states. 
In the triangular case, the clustering pattern
agrees with numerical results of Lecheminant {\it et al}
(Phys. Rev. B 52, 6647 (1995)).
\end{abstract}
\pacs{PACS numbers: 75.10.Jm, 03.65.Sq, 75.40.Mg, 75.30.Kz}

]

%
%
%
%

How can one identify the long-range order of an extended 
quantum system from exact diagonalizations that 
are severely constrained by finite-size effects? 
The commonest analysis is to extrapolate the 
ground-state correlations and energy gaps 
infinite size. 
An alternative approach uses
the finite-size splittings and symmetries of an entire family
of low-lying states to characterize possible 
symmetry breaking.~\cite{diagtri,lech95} 
In the latter spirit, we present a semiclassical calculation of 
the low-lying singlet eigenstates and eigenenergies
of four-sublattice Heisenberg antiferromagnets, 
assuming long-range order in the $N\to \infty$ limit
(where $N$ is the number of spins). 
The Hamiltonian is
   \be
      \Hop(\{\ss_i\}) =  {1\over 2} \sum _{ij} J(\rr_{ij}) \ss_i\cdot \ss_j
   \label {eq-Ham}
    \end{equation}
where spin $\ss_i$ has quantum spin length  $s$, and $\rr_{ij}$
is the vector connecting sites $i$ and $j$. 
By ``four-sublattice'' antiferromagnet, we mean one for which
a classical ground state is any state
with spins parallel within each of the four
(equivalent) sublattices, 
and the vector sum of their sublattice magnetizations zero. 
Important examples 
are the triangular antiferromagnet~\cite{lech95,chu92},  with
$J_1>0$ and second-neighbor coupling $J_2\in (J_1/8,J_1)$, 
and the Type I fcc antiferromagnet~\cite{og85,hen87} with 
$J_1>0, J_2 <0$. 
After allowing for rotations, there is still a two-parameter family 
of classical ground states.~\cite{chu92,og85,hen87}. 

These special degeneracies
will get split in a fashion characteristic of four-sublattice order, 
producing a characteristic pattern of low-lying eigenstates. 
To derive this pattern, 
we will map the system (approximately) into a 4-spin  one, 
and in turn to a sort of one-spin system, 
very much like the well-known cubic-symmetry molecular 
rotor~\cite{harter,robbins}. 
Using Bohr-Sommerfeld (BS) quantization of
classical orbits to define energies, 
which are split by tunneling between the classical orbits, 
we predict below a characteristic pattern of level clustering 
in accord with the numerical
data\cite{lech95} for the triangular case with $N=28$, the largest 
size studied to date. 

Let $\LC_\alpha$ be the set of $\Nsub\equiv N/4$ sites
in sublattice $\alpha$ ($\alpha=1,2,3,4$). 
The sublattice magnetizations are four big spins
$  \SS_\alpha \equiv \sum_{i\in \LC_\alpha} \ss_i$. 
Inspired by the classical ground states, we will define
``\fourspin'' states as those in which (i) each $\SS_\alpha$ has maximum 
total spin length $|\SS_\alpha| = \Ssub\equiv \Nsub s$; 
(ii) the total (quantum) spin $\ST \equiv \sum_\alpha \SS_\alpha$ 
is zero (singlet). 
This is the low-energy singlet subspace, 
as first recognized  by Lecheminant {\it et al}\cite{lech95,FN-lech}.
It has $2\Ssub+1$ independent quantum states.

Our first goal is a reduced Hamiltonian acting only
on \fourspin states. 
We first obtain an infinite-range version of (\ref{eq-Ham}) --
{\em i.e.}, every spin interacts equally with every spin of a different sublattice --
by replacing each $\ss_i \to\SS_\alpha/\Nsub$ in $\Hop$, where
$i \in \LC_\alpha$:
   \be
        \Hinf =
        {1\over 2} \tJ_0  \ST^2 
         - {1\over 2} \tJ \left(\sum _\alpha \SS_\alpha^2 \right)
    \label{eq-Hinf}
    \end{equation}
where $\tJ_0\equiv(3\Nsub)^{-1} \sum^{diff} _\rr J(\rr)$, 
$\tJ \equiv \tJ_0-
     {\Nsub}^{-1} \sum^{same}_\rr J(\rr)$. 
(Here ``$same$'' and ``$diff$''
mean the sums are restricted to interactions connecting the same
or different sublattices.) In the triangular case $\tJ=\tJ_0=2(J_1+J_2)/\Nsub$. 

The ``\fourspin'' states (having $\ST=0$ and maximum $|\SS_\alpha|$)
are manifestly the 
only ground states of $\Hinf$, having energy $\Eal = -2\tJ \Ssub(\Ssub+1)$.
Their degeneracy is broken 
by an effective Hamiltonian $\Hsel(\{\SS_\alpha\})$ which 
``selects'' particular ground states; 
this will be crudely approximated by a phenomenological biquadratic form
   \be
      \Hbiq \equiv
      - \tK \sum _{\alpha<\beta} (\SS_\alpha \cdot \SS_\beta)^2  + \tCbiq
   \label{eq-Hbiq}
   \end{equation}
with $\tK>0$.

Eq.~(\ref{eq-Hbiq}) can derived from
$\delta \Hop \equiv \Hop-\Hinf$ 
by either of two perturbative approaches. 
First, in the large-$s$ limit, and after the
usual Holstein-Primakoff expansion around a 
\fourspin coherent state, 
$\delta\Hop$ could be approximated by a harmonic Hamiltonian
which is diagonalized by spin-wave states. 
The spin-wave zero-point energy, summed over all modes
\cite{shender,hen89},
can then be expressed as a function of the four spin directions. 
By general (``order due to disorder'') arguments, 
collinear (parallel {\it or} antiparallel) spin
configurations
have the lowest zero-point energy~\cite{shender,hen89}, 
and Eq.~(\ref{eq-Hbiq}) is the simplest analytic form
with this property ~\cite{larson,jacobs}. 

Alternatively, Eq.~(\ref{eq-Hbiq}) can be directly
derived from second-order perturbation theory in $\delta \Hop$. 
This approximation~\cite{FN-2ndorder}
(related to that of \cite{larson})
gives precisely the form (\ref{eq-Hbiq}), with
    \be
     \tK = 
\frac{J_1^2+J_2^2- 2 \Nsub ^{-1} (J_1+J_2)^2 }
     {\Nsub \tJ  \Ssub (2\Ssub-1)^2 }
    \label{eq-tK}
    \end{equation}
for the triangular case, and
$\tCbiq= -2\Ssub^2 (5 \Ssub^2 - 6\Ssub-2) \tK$. 
We caution that the (presumably large) quantum fluctuations within each 
sublattice ought to renormalize (\ref{eq-tK}) by a substantial factor. 

From here on we work with the effective \fourspin Hamiltonian 
$\Eal+\Hbiq$. 
The quantum eigenstates of $\Hbiq$ will be approached semiclassically;
that is valid in a moderately large system even for $s=1/2$,  
since $\Ssub=\Nsub s$ is the large parameter in $\Hbiq$. 
Our classical coordinates are the unit vectors $\mm_\alpha$ that
parametrize the \fourspin coherent states, 
constrained (just like four-sublattice classical ground states)
only by $\sum_\alpha \mm_\alpha=0$. 
It will be useful to define coherent states 
$|\mmstate{}\rangle$ for the \fourspin problem, 
meaning each $\SS_\alpha$ has maximum projection 
in the direction $\mm_\alpha$.
There are competing prescriptions to derive the
effective classical Hamiltonian $\Ubiq$ from $\Hbiq$, which 
disagree after the leading power in $\Ssub$; we adopt~\cite{shankar,klauder}
    \be
     \langle \mmstate{}|\Hbiq|\mmstate{}\rangle 
    = \Ubiq(\mmstate{}) +C_U, 
     \label{eq-Hclassical}
     \end{equation} 
where
     \be
      \Ubiq (\mmstate{})=  
         -K_U \sum _{\alpha < \beta}   (\mm_\alpha \cdot \mm_\beta)^2
    \label{eq-Ubiq}
    \end{equation}
with $K_U= \Ssub^2(\Ssub-1/2)^2 \tK$ and 
$C_U = \Ssub^2(6\Ssub+5/2)\tK  + \tCbiq$. 

Now, the Green's function of such a system can be expressed as a 
path integral,\cite{shankar,klauder}
in which each path is weighted as usual by 
  \be
     \exp \left[ \int dt \Ubiq(\mmstate{}) - \Phi \right]
  \label{eq-pathint}
  \end{equation}
where the {\it topological phase} of the path is given by
   \be
      \Phi  = \sum _{\alpha=1}^{4} \Ssub \Omega(\{\mm_\alpha(t)\}), 
   \label{eq-topo}
  \end{equation}
where $\Omega(\{\mm(t)\})$ means the spherical area the
trajectory $\mm(t)$ has swept out on the unit sphere 
about its ``north pole''~\cite{FN-halves}. 
The classical dynamics is
   \be 
      d\mm_\alpha /dt = \gamma \mm_\alpha \times 
      \delta \Ubiq(\mmstate{})/d\mm_\alpha . 
    \label{eq-classdyn}
   \end{equation}

The dynamics in the 5-dimensional subspace $\sum \mm_\alpha =0$
is separable by a change of variables from
$\mmstate{}$ to a unit vector $\nn$ and a proper rotation matrix
$\RC$ defined as follows. 
Let $\nnmu \equiv \frac{1}{2}
(\mm_\mu +\mm_4)$, $\mu =1,2,3$ \cite{hen87};
these three vectors are orthogonal as 
follows from $\sum \mm_\alpha =0$. 
The proper rotation matrix $\RC$ is defined to align these vectors 
(in either sense) along the respective coordinate axes $\ee_\mu$, 
so $\nnmu\equiv n_\mu \ee_\mu$ and $\nn\equiv (n_1,n_2,n_3)$
is a unit vector. 
A discrete redundancy remains, that $\RC$ is well defined 
only up to a $\pi$ rotation of $\nn$ about any coordinate axis. 

We are finally interested in the {\it singlet projection}
of the coherent states, which is equivalent to averaging 
over all $\RC$ (with correct phase factors). 
This singlet basis is labeled only by $\nn$; it is important
for the sequel that $\nn$'s 
related by the redundancy  correspond to {\it identical}
singlet-projected basis states. 

Substituting from (\ref{eq-Ubiq}) shows
the Hamiltonian depends only on $\nn$:
    \be
       \Ubiq(\nn) =  -8 K_U \sum _{\mu=1}^3 (n_\mu)^4 + 2 K_U
    \label{eq-Ucub}
    \end{equation}
When (\ref{eq-classdyn}) is translated into the coordinates $\nn$ and $\RC$, 
it turns out that $d\RC/dt \equiv 0$ while 
$d\nn/dt = \gamma' \nn\times \delta \Ubiq/\delta \nn$, 
identical to the classical dynamics of {\it one}
spin with Hamiltonian (\ref{eq-Ucub}), i.e., 
a cubic anisotropy field; these
classical orbits follow contours of constant energy 
on the unit $\nn$ sphere, in the
sense indicated by the arrows in Fig.~\ref{fig-orbits}. 
Furthermore, each $\mm_\alpha(t)$ traces out the same shaped trajectory and
makes the same contribution to the sum (\ref{eq-topo}). 
Hence, the total topological phase $\Phi= 4\Ssub \Omega(\{\nn(t)\})$ 
is the same as that of 
{\it one} spin with length $\Sall \equiv 4 \Ssub = \Nall s$. 
In (only) this sense, we have mapped a four-spin to a one-spin problem.

{\it Bohr-Sommerfeld orbits and tunnel splittings ---}
We can identify three kinds of classical ground state 
with high symmetry, corresponding to the special points
indicated on the $\nn$ sphere in Fig.~\ref{fig-orbits}:
(i) The ``collinear'' ($C$) states, with all spin directions
$\mm_\alpha$ oriented along the same direction 
(two being parallel to it, and the other two antiparallel);
these are the ground states of $\Ubiq$. 
There are three C states (since there are three ways to group the spins 
$\{\mm_\alpha \}$ into pairs). 
On the unit $\nn$ sphere,  $C_{X,Y,Z}$ lie along the $\pm x$, $\pm y$
and $\pm z$ coordinate axes; these six points really correspond to
three distinct states in view of the discrete redundancy.
(ii) ``Tetrahedral'' ($T$) states
in which the spins point towards the corners of a
regular tetrahedron in spin space, the maximum energy states
of $\Ubiq$. 
There are two $T$ states since there is a  right-handed and a 
left-handed way of orienting the tetrahedron. 
These correspond to points $(\pm 1,\pm 1, \pm 1)/\sqrt{3}$ on the
$\nn$ sphere, corresponding to only two distinct states
after applying the discrete degeneracy. 
(iii) ``Saddle'' or ``square'' ($S$) states, in which the four spin directions
lie in the same plane in spin space and differ by $90^\circ$ rotations;
these are saddle-points of the $U_{biq}$ function. 
The corresponding values of $\Ubiq$ (using (\ref{eq-Ubiq}) are
$U_{C}=-6K_U$, 
$U_{S}=-2K_U$, and
$U_{T}=-{2\over 3} K_U$.

The \BS quantization condition~\cite{shankar}
selects orbits around $T$ with energies 
$\Ubiq(\nn)=U_{Tl}$ such that $\Phi = 2 \pi l$ for $l=0,1,2,\ldots$. 
However, the \BS condition for the energy $U_{Cl}$ of a $C$ type orbit is
that the complete loop has phase $\Phi= 4\pi l$;
this is due to the redundancy of $\nn$, 
whereby a $C$ type orbit returns to an equivalent state 
(and completes the true orbit) after looping just halfway around the C point. 
We will label each orbit by its \BS quantum number $l=0,1,\ldots$ and
by the label of the fixed point it encircles, thus ``$T_{l\gamma}$''
(for $\gamma=\pm$, a twofold degeneracy) 
or ``$C_{l\gamma}$'' (for $\gamma=X,Y,Z$, a threefold degeneracy.)
See Fig.~\ref{fig-orbits}. 

Each level cluster is built from \BS orbits degenerate in energy. 
Depending upon the topological phase~\cite{topo}, 
tunneling may occur between these orbits and slightly split this degeneracy. 
Let's write, e.g.,  $t_{Cl}(Y[c]X)$ to mean the 
total amplitude to ``hop'' from orbit $C_{lX}$ to $C_{lY}$ 
along paths which separate from $C_{lX}$, 
pass near saddle point $S_c$, and join onto $C_{lY}$. 
(In light of the discrete redundancy of $\nn$, 
there was just one tunneling path connecting a pair of $C$ orbits.)
All symmetry-related hoppings have the same magnitude $t_{Cl}$
but their phases depend on the ``gauge'' choice 
used in defining coherent states.

The hopping between the three \BS states is just like that between 
three atomic orbitals in a ring threaded by a flux. 
The eigenenergies are $U_{Cl}+2 |t_{Cl}|\cos((2\pi j-{\rm Re}\Phi_{Cl})/3)$, 
where $j$ is any integer, and the gauge-invariant 
${\rm Re} \Phi_{Cl}$ is the cyclic sum of the three phase angles. 
We will write ``$(\nu_1,\nu_2)$''
for the pattern of eigenvalues in this  cluster, 
meaning the (lower,higher) eigenvalues have degeneracies
$(\nu_1, \nu_2)$. respectively. 

So we just need to know
$t_{Cl}^3 e^{i{\rm Re}\Phi_{Cl}}$
$= t_{Cl}(Z[a]Y)t_{Cl}(Y[c]X)t_{Cl}(X[b]Z)$, 
the amplitude for closed paths around the loop 
shown in Fig.~\ref{fig-orbits}(b). 
The stationary-phase trajectory, as shown in Fig.~\ref{fig-orbits}(b), 
follows orbit $C_{lX}$ to a point 
along the ``equator'',  then crosses the classically forbidden 
barrier by following a trajectory  
with part {\it complex} coordinates and joins onto orbit $C_{lY}$ 
such that $\Ubiq \equiv U_{Cl}$ along the entire path\cite{klauder}.
Because the classically forbidden segments 
follow high symmetry lines, the real part of the spherical area enclosed
(hence of $\Phi_{Cl}$) is computed from the real part of the trajectory
(shaded in the figure), and correspondingly for the imaginary parts. 
For example, the segment of stationary-phase path 
crossing $S_c$ in Fig.~\ref{fig-orbits}
has $\theta(\phi)=\pi/2 -i{\rm Im}\theta(\phi)$;
in the spherical area
$\Omega \equiv \oint d\phi(1-\cos\theta(\phi))$, 
the integrand for this segment has
$\cos \theta \to i \sinh ({\rm Im} \theta)$.

Thus
${\rm Re} \Phi_{Cl}$ is $\Sall$ times the spherical area enclosed by 
the loop shown in Fig.~\ref{fig-orbits}(b):
{\it i.e.} $\Sall (4\pi/8)-3(4\pi l/4)$, where the
second term comes from the arcs truncating the triangle corners 
in Fig.~\ref{fig-orbits} (b). 
Thus $e^{i{\rm Re}\Phi_{Cl}} = (-1)^{\Sall/2-3l}$. Recalling that
$\Sall \equiv \Nall s$, 
we obtain a cluster pattern $(2,1)$ if $\Nall s/2-3l$ is even 
or $(1,2)$ if it is odd.
On the other hand, 
${\rm Im} \Phi_{Cl}$ is just the WKB 
exponent appearing in the tunnel amplitude, thus
$t_{Cl} \sim e^{-{{\rm Im} \Phi_{Cl}/3}}$. 

The case of $T$ type orbits differs in that
there are three distinct stationary-phase paths connecting orbits
$T_{l+}$ and $T_{l-}$.
So, from the truncated-square loop shown in Fig.~\ref{fig-orbits}(c), we
find $t_{Tl}(+[b]-)^2 t_{Tl}(-[c]+)^2  = t_{Tl}^4 e^{i{\rm Re}\Phi_{Tl}}$, 
with ${\rm Re} \Phi_{Tl} = \Sall (4\pi/6) - 4(2\pi l/3)$. From 
the other loops related to it by symmetry, we get formulas
with $[b]...[c] \to [c]...[a] \to [a]...[b]$ as well, 
which give all the relative
phases between the three different paths connecting orbit
$T_{l+}$ to $T_{l-}$. The total hopping is a sum over these paths, 
$\ttot_{Tl} \equiv \sum _{p=a,b,c} t_{Tl}(-[p]+)$, and the 
relative phases of these terms turn out to be $0, \pm {\rm Re} \Phi_{Tl}$. 
So when $\Nall s
-4l$ is divisible by 3, then $\ttot_{Tl}\neq 0$ and
the cluster pattern is $(1,1)$ with splitting $2|\ttot_{Tl}|$;
otherwise, $\ttot_{Tl}=0$ 
and the cluster pattern is $(2)$ (i.e., unsplit)~\cite{FN-threefold}.

The overall clustering pattern thus begins (starting with the ground state) 
with repeats of 
$(2,1)(1,2)\ldots$ for $C$ orbits -- for odd $\Nall s /2$, the first cluster
is $(1,2)$ -- and turns into repeats of $\ldots 2(1,1)2 \ldots$ for
the $T$ orbits corresponding to the higher energies among the
low-lying singlets. Amusingly, the sequence of degeneracies $\{2,1,1,2\}$
continues unbroken past the saddle-point energy $U_S$, so there is
no sharp boundary between the two behaviors. 

{\it Comparison to diagonalizations ---}
The spin-$1/2$ triangular system was exactly 
diagonalized by Ref.~\onlinecite{lech95}. 
Table~\ref{tab-levels} shows 
the numerical data of Lecheminant {\it et al} \cite{lech95} 
for the $s=1/2$ triangular antiferromagnet
with $(J_1, J_2)= (1, 0.7),$ 
for $\Nall =16$ and $\Nall =28$. 
All of their low-lying singlet energies are given, compared with
our numerical predictions, as $\Delta E$ (the difference $E-\Eal$). 
Each row is one cluster, labeled in column 1 with the 
\BS orbit from which it derives, and in parentheses
the degeneracies of the levels in the cluster
(from lowest to highest). 
The other columns give the mean energy of the cluster, and
in brackets its tunnel splitting $t_{Cl}$ or $\ttot_{Tl}$ (if nonzero). 
The ``4-spin'' column is (\ref{eq-Hbiq}) with 
$\tK$ given by (\ref{eq-tK});
the ``Theory'' column is (\ref{eq-Hclassical}) and (\ref{eq-Ucub}) 
with $K_U$ given after (\ref{eq-Ubiq}). 
(In the absence of a theory for their prefactor, 
we estimated the ``theory'' splittings as
$t_{Cl} \to (8K_U)\exp(-\Sall \;{\rm Im}\Phi_{Cl}/3)$, 
where ${\rm Im}\Phi_{Cl}/3 =0.55$ for $l=0$ and $0.10$ 
for $l=1$.)

The eigenvalue clusters 
(which comprise all the low-energy states in these small systems)
fall in exactly the pattern we predict. 
For $\Nall=16$, there is a near-cancellation in formula (\ref{eq-tK});
this explains qualitatively why the whole energy scale is so small
compared to $\Nall=28$, 
and why quantitative agreement between theory and experiment is not
expected in the $N=16$ case.
For $\Nall=28$, 
the cluster energies agree fairly well, apart from a constant offset
which we cannot at present explain. 

{\it Summary and discussion --}
To conclude, we have identified two small energy scales
among the low-lying singlet states 
in four-sublattice antiferromagnets, also indexing all of these eigenstates
and explaining the observed pattern \cite{lech95} of their energy splittings. 
The smallest splittings 
(of $O(\exp (-{\rm const} \Nall s)) $)
are explained by tunneling between different classical wells that
result from discrete symmetry breakings; 
exact degeneracies occur when tunnel amplitudes summed
along multiple paths cancel, 
owing to the topological phase.~\cite{topo}
The next smallest splittings between singlets in the $N\leq 28$ systems 
are between clusters, i.e. between successive Bohr-Sommerfeld orbits. 
This spacing is proportional to the zero-point energy per spin
(favoring spin collinearity) which scales as a constant as $N\to\infty$. 

We can, of course, predict cases which have not yet been diagonalized:
{\it e.g.,} for $s=1/2$ in $\Nall =32$ (fcc case) or $\Nall=36$ (next larger
triangular system), 
the cluster patterns are respectively 
$(1,2)(2,1)1(2)$ and $(2,1)(1,2)(2)(1,1)$. 

For sufficiently large $N$, our assumption that spins in 
each sublattice stay rigidly aligned
must fail during the tunnel event:
the tunneling barrier will be smallest for an inhomogeneous
quantum-nucleates and then grows. 

Finally, it may be noted that our semiclassical treatment is 
exactly the same as that of a one-spin Hamiltonian with total
angular momentum $S$ and cubic anisotropy -- 
exactly the Hamiltonian analyzed semiclassically~\cite{harter,robbins}
to account for the rotational spectroscopy of $\rm SF_6$. 
The eigenstate clusters in Table~\ref{tab-levels}
are a subset of those in the one-spin system, 
as $3/4$ of the latter have symmetries forbidden in our system
(in view of the discrete redundancy of states labeled by $\nn$).

We are grateful to P.~Houle, K.~Nakamura, and E.~Heller for discussions,
and to the authors of Ref. [2] for comments and for supplying their
numerical data.
This work was supported by NSF grant 
DMR-9612304.

\begin{table}
\caption{Eigenenergies $\Delta E$ for small systems}
\label{tab-levels}
\begin{tabular}{llll}
Orbit & Exact\cite{lech95}  & 4-spin  &  Theory \\
\tableline
       &  $N=16$         &  $\Ssub=2$   &   $S=8$ \\
\tableline
$T_0$ (2)   & -0.151     &     -0.073 &       -0.0029 \\
$C_0$ (1,2) & -0.272(0.006) &  -0.137 (0.0029)    & -0.0323 (0.00065)\\
\tableline
       &  $N=28$         &  $\Ssub=7/2$   & $S=14$ \\
\tableline
$T_0$ (2)   &  -3.754    &        -1.919 &  -1.121 \\
$C_1$ (1,2) &  -4.021 (0.017)  &  -2.336 (0.037) & -1.478 (0.34)\\
$C_0$ (2,1) &  -4.769 (0.0003) &  -3.158 (0.0018) & -2.033 (0.00062)\\
\end{tabular}
\end{table}

\begin{figure}
{\epsfxsize=3.6truein\epsfbox{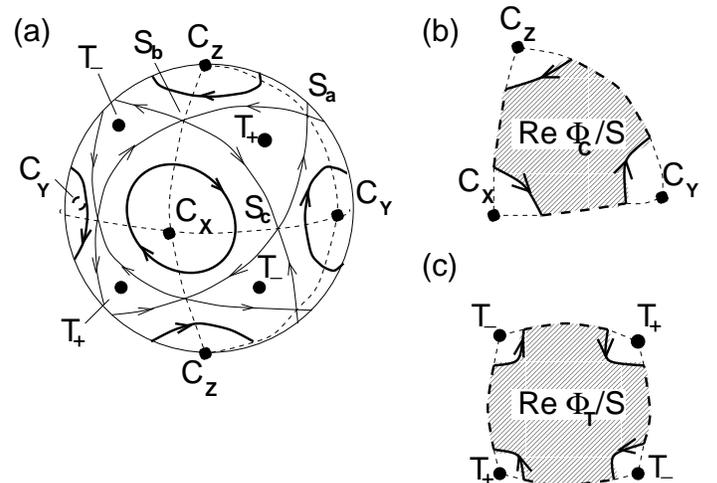}}
\caption{
Unit sphere representing a spin with classical 
orbits following contours of the Hamiltonian (\ref{eq-Ucub}).
Orbits $C_{0\alpha}$ ($\alpha=x,y,z$) and $T_{0\pm}$ 
sit on the special symmetry points $C_\alpha$ and $T_\pm$
(indicated by dots); orbits $C_{1\alpha}$ are shown by heavy lines, and 
separatrices connect the saddle points $S_a, S_b, S_c$. 
(The unmarked saddle points are equivalent to these by the
redundancy mentioned in text.)
(b). Loop for estimating tunneling amplitude $t_{Cl}$; the shaded
spherical area is indicated. Heavy dashed portions are classically forbidden
and contribute to ${\rm Im}(\Phi_C)$.
(c). Same but for $t_{Tl}$ type orbits. (This area has reversed sign since
this loop runs clockwise.)}
\label{fig-orbits}
\end{figure}


\begin{references}
\bibitem{FN-lech}
It was not checked whether their low-lying states are literally
\fourspin states, or merely connected adiabatically in
the spirit of Fermi liquid quasiparticle states.
\bibitem{FN-2ndorder}
The first key observation for this derivation concerns the matrix elements 
among states $|M\rangle$ of the multiplet with maximum
total $|\SS_\alpha|$, 
such as $\langle M'| s_{i\lambda} s_{j\mu}|M\rangle$, 
where $i, j \in \LC_\alpha$ with $i\neq j$
and $\lambda, \mu$ are Cartesian indices. 
Since $|M\rangle$ is completely symmetric under permutations of sites $i$, 
any such matrix element is independent of $\rr_{ij}$. 
The second key observation is to consider the action on
a ``\fourspin'' state of any term in $\delta \Hop$, 
say $(\ss_i-\SS_\alpha/\Nsub) \cdot (\ss_j - \SS_\beta/\Nsub)$
with $i\in \LC_\alpha$, $j\in \LC_\beta$ (with $\alpha\neq \beta$). 
It leaves $\ST=0$ but reduces $|\SS_{\alpha}|$
and $|\SS_{\beta}|$ to $\Ssub-1$. 
Hence the second-order energy denominators are all $2 \tJ S $.
\bibitem{FN-halves}
A \BS prescription with $S \to (S+1/2)$ in (\ref{eq-topo})
and a quantization condition $\Phi_{orbit} = 2\pi (l + 1/2)$
is empirically more accurate; see \cite{robbins}.
\end{references}
\end{document}